\documentclass[aps,prl,twocolumn]{revtex4}
\usepackage{amsmath}
\usepackage{graphics}
\usepackage{graphicx}
\usepackage{color}

\begin{document}

\title{Irreversible Magnetization Deep in the Vortex-Liquid State of a 2D
Superconductor at High Magnetic Fields}
\author{T. Maniv,$^{1,3}$ V. Zhuravlev,$^{1}$ J. Wosnitza$^{2}$, and J. Hagel$^{2}$}
\affiliation{$^1$Chemistry Department, Technion-Israel Institute of
Technology, Haifa 32000, Israel \\
$^2$Institut f\"ur Festk\"orperphysik, Technische Universit\"at Dresden,
D-01062 Dresden, Germany\\
$^3$Grenoble High Magnetic Field Laboratory, Max-Planck-Institute fur 
Festkorperforschung and CNRS, Grenoble, France}
\date{\today}

\begin{abstract}
The remarkable phenomenon of weak magnetization hysteresis loops, observed
recently deep in the vortex-liquid state of a nearly two-dimensional (2D)
superconductor at low temperatures, is shown to reflect the existence of an
unusual vortex-liquid state, consisting of collectively pinned crystallites
of easily sliding vortex chains.
\end{abstract}

\maketitle

%% \email{maniv@tx.technion.ac.il}

%%%%%%%%%%%%%%%%%%%%%%%%%%%%%%

Many potentially important superconductors, such as some of the high-$T_{c}$
cuprates, as well as the ET-based organic conductors, are extremely type-II
superconductors with small in-plane coherence lengths and nearly 2D
electronic structure. Thereby, ET (or BEDT-TTF) stands for
bisethylenedithio-tetrathiafulvalene. Drastic deviations from the
predictions of the mean-field theory for these materials, due to strong
fluctuations in the superconducting (SC) order parameter, are therefore
expected, in particular under strong perpendicular magnetic field \cite
{RMP01}. The great interest, from a fundamental point of view, in the latter
class of materials stems from their relatively low upper critical fields,
which enable to investigate the virtually unexplored phase diagram and
vortex dynamics of strongly type-II superconductors at low temperatures and
high magnetic fields.

\begin{figure}[b]
\begin{center}
\includegraphics[width=8.5cm]{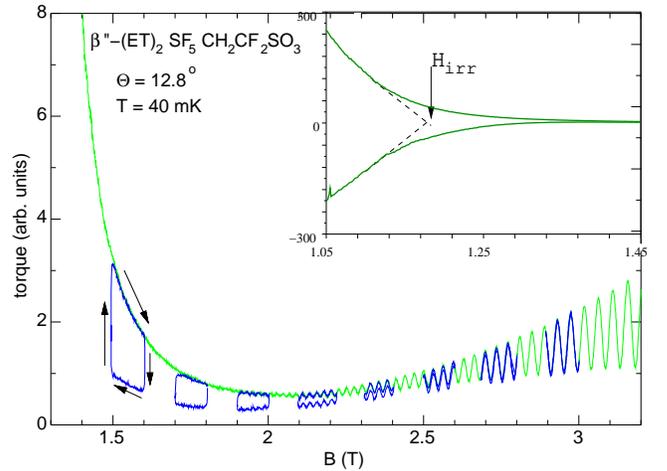}
\end{center}
\caption{Field dependence of the torque signal with several small hysteresis
loops in the tail of the hysteresis loop shown in the inset. Arrows indicate
the field-sweep directions. Inset: Lower field torque data showing how the
major irreversibility field is determined.}
\label{Fig.1}
\end{figure}

Of special interest is a recent striking observation in $\beta ^{\prime
\prime }$-(ET)$_{2}$SF$_{5}$CH$_{2}$CF$_{2}$SO$_{3}$ of small, but very
clear magnetization hysteresis loops appearing well above the `major'
irreversibility field at low temperature, where significant de Haas--van
Alphen (dHvA) oscillations are observable as well \cite{Wosnitza03} (Fig.\ 
\ref{Fig.1}). It should be stressed that the occurence of such high field
hysteresis tail is not peculiar to this particular material. It can also be
observed at low temperatures in $\kappa $-(ET)$_{2}$Cu(SCN)$_{2}$ \cite
{Sasaki98}.

The unusual feature of this irreversibility effect is associated with its
appearance deep in the vortex-liquid phase, where one usually expects
unrestricted motion of flux lines through the entire SC sample, ensuring the
establishment of thermodynamic equilibrium. In the present letter we argue
that the observed hysteresis effect is a general feature which reflects the
unusual nature (namely the nematic liquid-crystalline structure) of the
low-temperature vortex-liquid state in 2D superconductors well above the
vortex-lattice melting point, which was recently predicted theoretically 
\cite{ZM02}.

According to this model the vortex state above the major irreversibility
field, $H_{irr}$, (which is found to be close to the vortex-lattice melting
field, see below) is not an isotropic liquid but some mixed phase containing
SC domains of easily sliding parallel vortex chains, which are stabilized by
a small number of strong pinning centers. This model resembles the sluggish
vortex fluid picture, proposed by Worthington et al \cite{Worthington92} to
describe the intermediate vortex liquid phase in defect-enhanced YBCO singel
crystals, and applied more recently by Sasaki et al \cite{Sasaki02} to
account for a similar behavior in the organic superconductor $\kappa $-(ET)$%
_{2}$Cu(SCN)$_{2}$ . As shown below, a simple Bean-like model for the
magnetic-induction profiles, associated with the injected vortex chains into
the SC sample, yields very good quantitative agreement with the measured
field dependence of the magnetization hysteresis.

\begin{figure}[bp]
\begin{center}
\includegraphics[width=2.5in]{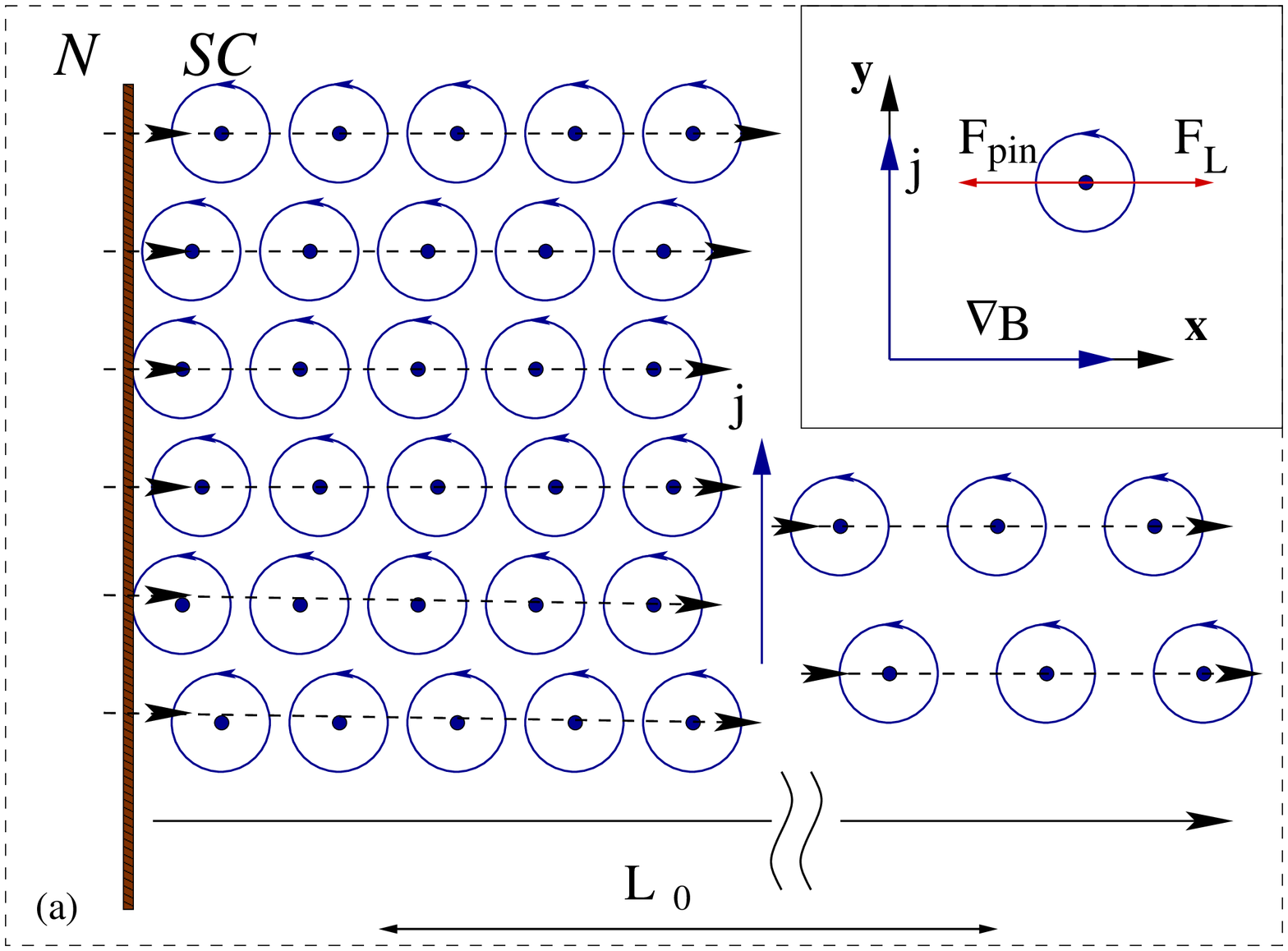} \\[0pt]
\includegraphics[width=2.5in]{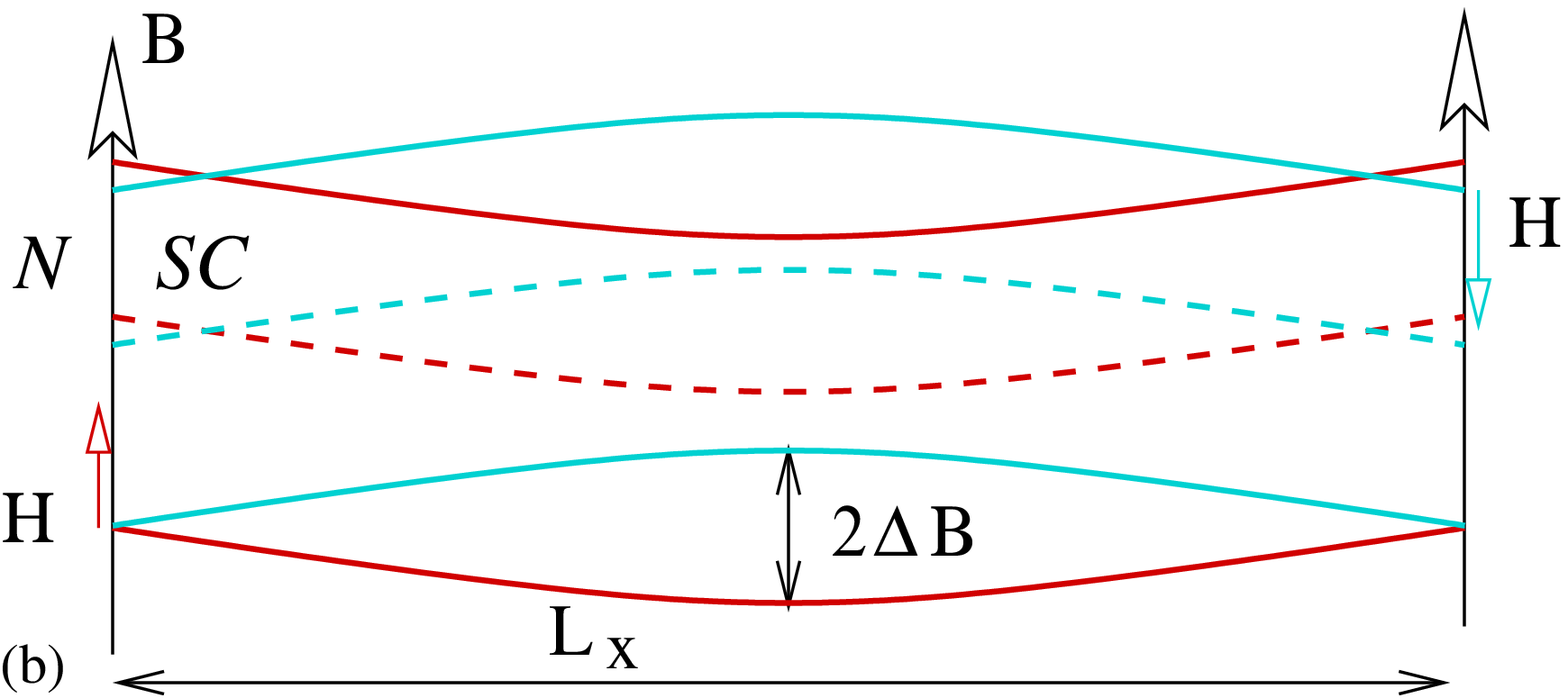}
\end{center}
\caption{(a) Sketch of a vortex crystallite, oriented with its easily
sliding Bragg chains perpendicular to the N-S boundary, under injection of
flux lines. Note the scaling of the magnetic length over a macroscopic
distance $L_{0}$. (b) Magnetization profiles during upward and downward
field sweeps.}
\label{Fig.2}
\end{figure}

Our model consists of independent 2D SC layers in the $x$-$y$ plane under
the influence of an external uniform magnetic field $\vec{H}=H\widehat{z}$, $%
H>0$, perpendicular to the layers (Fig.\ \ref{Fig.2}). The SC sample is in
contact with a normal metal at two parallel planes, $x=0$, and $x=L_{x}$. $H$
is varied, at low temperatures, from above $H_{c2}$ to the irreversibility
field $H_{irr}<H_{c2}$, where magnetic quantum oscillations can be detected.
The dHvA effect is measured by means of the torque method, in which the
signal is detected during steady (upward and downward) sweeps of the
external magnetic field. Note that, for the sake of simplicity, the small
in-plane component of the external magnetic field, required for this
high-resolution measurement, is neglected in our analysis.

The influence of a steady increase (or decrease) of the magnetic field on
the distribution of magnetic flux lines within the SC region is determined
by the pinning forces acting on quantized SC vortex lines near the
normal-superconducting (N-S) boundary planes. The pinning-force resistance
against flux injection leads to the establishment of a flux-density gradient
perpendicular to the magnetic field direction along the normal of the N-S
boundary plane (along $x$). Assuming the flux injection to be uniform along $%
y$, the current density $\vec{j}=j\widehat{y}$ and the field gradient, $%
\frac{\partial B} {\partial x}$, are connected by Ampere's law: 
\begin{equation}
\frac{\partial B}{\partial x}=-\left[ \vec{\nabla }\times \vec{B}\right]
\cdot \widehat{y}=-\frac{4\pi }{c}j,  \label{Ampere}
\end{equation}
where the Lorentz-force density (per unit volume), exerted on the vortex
current by the magnetic field, is 
\begin{equation}
\vec{F}_L = \frac{1}{c}\left[ \vec{j}\times \vec{B}\right] = F_L\widehat{x}.
\label{Lorenz}
\end{equation}

For magnetic fields below $H_{irr}$, the pinning force, $F_{pin}=-\frac{%
\partial U_{pin}}{\partial x}$, is sufficiently strong to balance the action
of the driving Lorentz force, $F_{L}=\frac{1}{c}jB=-\frac{\partial U_{L}}{%
\partial x}$, with $U_{L}(x)=\frac{1}{8\pi }\left[ B\left( x\right) \right]
^{2}$, so that $\frac{\partial U_{tot}}{\partial x}=0$, with $U_{tot}\left(
x\right) \equiv U_{pin}\left( x\right) +U_{L}(x)$. Consequently, the static
magnetic-field gradient, $\frac{\partial B}{\partial x}$, established across
the sample, is followed by a dissipationless current parallel to the
surface. This static equilibrium situation persists under the increasing
local magnetic potential $U_{L}(x)$ near the surface, up to the `elastic
limit' of the pinning potential, $U_{pin}(x)$, above which bundles of
vortices, accumulated near the N-S boundary, can jump over (or tunnel
through) the $U_{tot}$-potential barriers to neighboring potential wells.
Such collective vortex jumps into the SC region quickly reduce the local
Lorentz force by relieving the large local field gradient, so that the
movement stops after a few steps. The observed sharp magnetization jumps in
this field range \cite{Wosnitza03,mola01} seem to be associated with this
effect.

The vortex-vortex correlations in this field range are relatively strong and
long ranged, so that the effective pinning force is long ranged as well. The
effect is particularly significant at low pinning-center density, where
these correlations can dramatically enhance collective-pinning processes,
since many vortices are involved in the interaction with a single pinning
center. Furthermore, due to the strong anisotropy of the vortex-vortex
coupling \cite{ZM02}, this collective-pinning mechanism is also very
anisotropic. Thus, the magnetic-flux injection should encounter different
resistances at different locations of the N-S boundary, depending on the
local orientation of the vortex crystallite with respect to the surface. The
most favorable orientation is that of crystallites having their principal
lattice vector aligned perpendicular to the N-S boundary, i.e., along $x$,
where a weak driving force can inject many vortex chains into the SC region.

As the external magnetic field approaches $H_{irr}$ from below and the
strength of the pinning potential is decreasing (because $U_{pin}$ is
proportional to the local superfluid density, which decreases with
increasing $B$), the increasing Lorentz force can exceed the maximal value
of the pinning force, i.e., the highest maximum of $U_{tot}(x)$ turns into
an inflection point. Above this critical field, $U_{tot}(x)$ has no barriers
and the motion of vortices in the entire SC region becomes a continuous
flow. This vortex movement with velocity $\vec{v}_{\phi }$ is opposed by a
dynamic friction force, $\vec{F}_{\eta }=-\eta \vec{v}_{\phi }=F_{\eta }%
\widehat{x}$, which balances the action of the driving Lorentz force at the
critical velocity $v_{\phi }^{c}$: 
\begin{equation}
F_{L}^{c}\equiv \frac{1}{c}j_{c}B=\left| \vec{F}_{\eta }\right| =\eta
v_{\phi }^{c}.  \label{CritLorenz}
\end{equation}
In the corresponding steady state the motion of flux lines generates an
electric field \cite{AbrikBook} 
\begin{equation}
\vec{E}=\frac{1}{c}\left[ \vec{B}\times \vec{v}_{\phi }^{c}\right] =E%
\widehat{y},  \label{Relative}
\end{equation}
with $E=\frac{1}{c}Bv_{\phi }^{c}$ parallel to the current density $\vec{j}$%
. Consequently, the system develops a finite electrical resistivity to the
critical-current flow, $j_{c}=c\eta v_{\phi }^{c}/B$, namely $\rho
(B)=E/j=B^{2}/c^{2}\eta $. Now, the Bardeen-Stephen (BS) relation \cite
{BS65,Blatter94}, $\eta \approx BH_{c2}/c^{2}\rho _{n}$ , which is
equivalent to the simple linear magnetoresistivity \cite{AbrikBook} $\rho
=(B/H_{c2})\rho _{n}$, together with Eqs.\ (\ref{Ampere}) and (\ref{Relative}%
), lead to: 
\begin{equation}
\frac{\partial B}{\partial x}=-\left( \frac{4\pi H_{c2}}{c^{2}\rho _{n}}%
\right) v_{\phi }^{c}.  \label{CritGrad}
\end{equation}

The Faraday induction law together with Eq.\ (\ref{Relative}) yield the
flux-line conservation law, $\frac{\partial B}{\partial t}+\frac{\partial
(v_{\phi }B)}{\partial x}=0$, which may be combined with Eq.\ (\ref{CritGrad}%
) to obtain the nonlinear `diffusion' equation for the magnetic-flux
density: 
\begin{equation}
\frac{\partial B}{\partial t}=\frac{D_{\phi }}{H_{c2}}\left[ B\left( \frac{%
\partial ^{2}B}{\partial x^{2}}\right) +\left( \frac{\partial B}{\partial x}%
\right) ^{2}\right],  \label{DiffEq}
\end{equation}
with the diffusion coefficient $D_\phi\equiv \frac{c^2\rho_n}{4\pi}$. In the
high-field limit of interest here, when the range $\Delta H$ of a single
field-sweep cycle is much smaller than the initial field of the hysteresis
loop, $H_{0}$, the nonuniform part of the magnetic induction associated with
the flux flow is very small, i.e., $\left| b(x,t)\right| \equiv \left|
B(x,t)-H(t)\right| \ll H(t)\simeq H_{0}$. Under this condition the second
term on the RHS in Eq.\ (\ref{DiffEq}) can be neglected, and the equation
may be linearized to $\frac{\partial B}{\partial t}\approx \left( \frac{H_{0}%
}{H_{c2}}\right) D_{\phi }\frac{\partial ^{2}B}{\partial x^{2}}$. The
solution of this equation, satisfying steady upward and downward field-sweep
boundary conditions, $\frac{\partial B}{\partial t}(x=0,t)=\frac{\partial B}{%
\partial t}(x=L_{x},t)=\frac{\Delta H}{\tau }$ for $0\leq t\leq \tau $, and $%
-\frac{\Delta H}{\tau }$ for $\tau <t\leq 2\tau $, takes the forms: $%
B_{+}(x,t)=H_{0}+\frac{\Delta Ht}{\tau }+\Delta B\left[ \left( \frac{2x}{%
L_{x}}-1\right) ^{2}-1\right] $, and $B_{-}(x,t)=H_{0}+\Delta H\left( 2-%
\frac{t}{\tau }\right) -\Delta B\left[ \left( \frac{2x}{L_{x}}-1\right)
^{2}-1\right] $, respectively, with $\tau $ half the period of the
field-sweep cycle.

Note the spatial rigidity of the induction profiles $B_{\pm}(x,t)$, which is
assured, in the high-field limit $\left| b(x,t)\right| \ll H$, by the nearly
instantaneous propagation of any local fluctuation of the magnetic-flux
density. Indeed, the propagation velocity of such a fluctuation, $v_f=\frac{%
\partial B}{\partial t}/\frac{\partial B} {\partial x}$, is found by Eq.\ (%
\ref{DiffEq}) to be: $v_f=\frac{c^2\rho_n} {4\pi H_{c2}}\frac{\partial }{%
\partial x}\left( B\frac{\partial B} {\partial x}\right) /\frac{\partial B}{%
\partial x}$, so that the flux-line velocity $v_\phi$ (see Eq.\ (\ref
{CritGrad})) satisfies: $\left| v_\phi\right| \simeq \left| v_{f}\right|
\left( \frac{\partial B} {\partial x}\right)^2/B\left| \frac{\partial^2B}{%
\partial x^2}\right| \ll |v_f|$.

\begin{figure}[tbp]
\begin{center}
\includegraphics[width=3.4in]{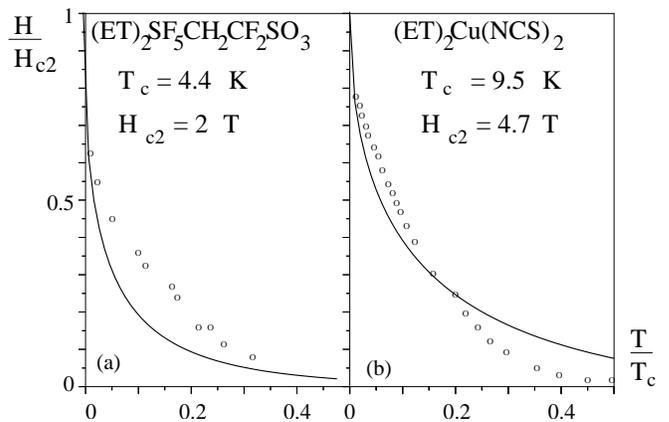}
\end{center}
\caption{Experimental major irreversibility fields (open circles), and
theoretical melting fields (solid lines), as functions of temperature for
two organic superconductors. Data for $\protect\kappa$-(ET)$_2$Cu(NCS)$_2$
from Ref.\ \protect\cite{Sasaki98}.}
\label{fig:4}
\end{figure}

The resulting Bean-like induction profile can now be exploited to evaluate,
for each small hysteresis loop, the jump, $\Delta M_{\uparrow \downarrow }$,
of the magnetization occurring at a point where the field sweep is reversed
(Fig.\ \ref{Fig.2}). The corresponding jump, $\Delta M_{\uparrow \downarrow
}=\frac{\zeta }{3\pi }\Delta B$, is determined by the spatial average of the
induction difference, $\left[ B_{-}\left( x,\tau \right) -B_{+}\left( x,\tau
\right) \right] $, where the (field-sweep rate-dependent) parameter $\zeta$ (%
$\zeta \leq 1$) is an effective volume fraction of the SC domains in the
sample. This simple Bean-like model may be tested by estimating the
parameter $\rho_n$, which should be identified with the in-plane resistivity
at $H_{c2}$ of the 2D conductor under study. This can be done by using the
experimental values: $4\pi \Delta M_{\uparrow \downarrow }\approx 4\times
10^{-7}$~T (see below), $L_x\approx 0.1$~cm, and $\left( \frac{\Delta H}{%
\tau }\right) \approx 0.005$~T/s \cite{Wosnitza03}, in the relation $(\Delta
H/\tau)\zeta =8D_\phi(H_0/H_{c2})\Delta B/L_x^2$, which yields $%
\rho_n\approx \frac{\zeta L_x^2}{3c^{2}\Delta M_{\uparrow \downarrow }}%
\left( \frac{\Delta H}{\tau }\right) \approx 0.4\zeta \times 10^{-6}\,\Omega 
$cm. Assuming that $\zeta $ is of the order unity, this is a reasonable
result, (i.e., of the order of a typical metallic resistivity), which cannot
be easily verified experimentally, since for the very anisotropic metal
under study here, measurements of the in-plane resistivity are extremely
sensitive to out-of-plane scattering by a minute amount of defects.

It is interesting to estimate the characteristic energy scale corresponding
to the friction of moving vortices in the liquid state above $H_{irr}$
against the effective pinning force. This can be done by estimating the
total spatial variation of the magnetic induction within the SC sample, $%
L_{x}\overline{\left| \frac{\partial B}{\partial x}\right| }\approx \zeta
\Delta B$, with the bar indicating spatial averaging, and comparing it to
the experimental value of $\Delta M_{\uparrow \downarrow }$. Thus, the work
(per unit volume) done by the driving Lorentz force against the friction
force, $E_{fric}=\frac{1}{c}j_{c}BL_{x}=\frac{1}{4\pi }B\overline{\left| 
\frac{\partial B}{\partial x}\right| }L_{x}\approx \frac{\zeta }{4\pi }%
B\Delta B$, may be rewritten in units of the maximal SC condensation-energy
density, $E_{cond}=\frac{H_{c2}^{2}}{16\pi \kappa ^{2}}$, that is for $%
B\approx H_{c2}$, $\widetilde{\varepsilon }_{fric}\equiv
E_{fric}/E_{cond}\approx 4\kappa ^{2}\zeta \Delta B/H_{c2}$. Using the
experimental value of the magnetization jump at 2~T, we find $\zeta \Delta
B=3\pi \Delta M_{\uparrow \downarrow }\approx 4\times 10^{-7}$ T, so that $%
\widetilde{\varepsilon }_{fric}\approx 1.7\times 10^{-3}$, where $\kappa
\approx 46$ is used \cite{Wanka98}. This estimate is smaller then the
characteristic energy associated with the melting of the vortex lattice in
2D superconductors, which is about $2\%$ of the SC condensation energy. It
may indicate that most of the vortex crystallites are aligned with their
principal lattice vector along the direction of the flux-line motion \cite
{Reichhardt00}, so that the flux flow is dominated by easily sliding vortex
chains, which are subject only to the residual shear resistance
characterizing the vortex-liquid state above the melting transition \cite
{RMP01}. This picture is very similar to the moving smectic state found
recently in many numerical simulations of 2D vortex lattices in the presence
of random pinning centers \cite{Olson98}.

\begin{figure}[tbp]
\begin{center}
\includegraphics[width=2.5in,angle=-90]{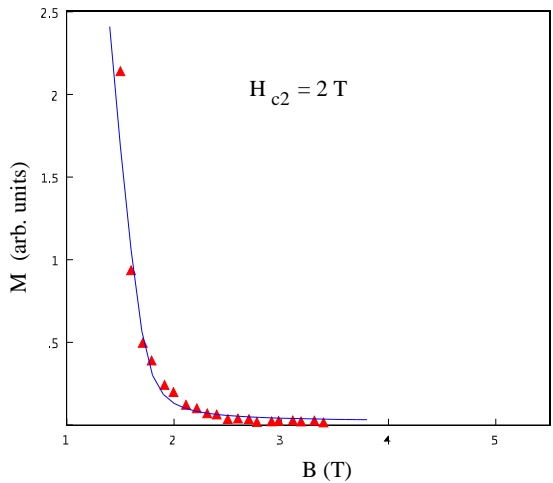}
\end{center}
\caption{Field dependences of the experimental magnetization jumps
(triangles) and the fit (solid line) according to Eq.\ (\ref{DelMjump}).}
\label{fig:3}
\end{figure}

The value of $\widetilde{\varepsilon }_{fric}$ estimated above may also
reflect the correlation between the onset of irreversibility and the
vortex-liquid freezing transition. This may be verified by calculating the
temperature-dependent melting field, $H_{m}(T)$, and comparing it to the
irreversible field, $H_{irr}(T)$, extracted from the experimental data.
Similar to Ref.\ \cite{Sasaki98}, $H_{irr}(T)$ is determined here by the
onset of the major hysteresis in the magnetization curve ( see Fig. \ref
{Fig.1} ). Within the Ginzburg--Landau (GL) approach, $H_{m}(T)$ has been
derived by several authors \cite{tesanovic94,ZM02}. It can be obtained from $%
\xi ^{2}(H_{m},T)=g_{m}^{2}$, where $\xi ^{2}(H,T)\equiv \varepsilon
_{0}/k_{B}T$, with $\varepsilon _{0}=\alpha ^{2}/2\beta $ being the SC
condensation energy per unit vortex, and $\alpha $ and $\beta $ are the GL
parameters \cite{RMP01}. Here we estimate $|\Delta _{0}|^{2}=\frac{\alpha }{%
\beta }=\left( 1.76k_{B}T_{c}\right) ^{2}\ln \left( \frac{H_{c2}}{B}\right) $%
, and $\beta =\frac{1.38}{E_{F}(\hbar \omega _{c})^{2}}$. Using our
estimate, $g_{m}^{2}\approx 1/4\lambda ^{2}$, with $\lambda \simeq 0.066$
the effective minimal shear energy relative to $\varepsilon _{0}$ \cite{ZM02}%
, and neglecting the weak temperature dependence of $H_{c2}$ at low $T$, we
obtain the simple equation for $h_{m}(t)=H_{m}(T)/H_{c2}(0)$: $\ln \left(
h_{m}\right) =-k_{0}\sqrt{t}h_{m}$, where $t=T/T_{c}(0)$ and $k_{0}$ is a
single dimensionless parameter depending on the properties of the
superconductor through $k_{0}^{2}\simeq 0.15g_{m}^{2}E_{F}(\hbar \omega
_{c2})^{2}/(k_{B}T_{c})^{3}$. The function $h_{m}(t)$ for $\beta ^{\prime
\prime }$-(ET)$_{2}$SF$_{5}$CH$_{2}$CF$_{2}$SO$_{3}$, with $T_{c}(0)=4.4$~K
and $H_{c2}=2$~T is shown in Fig.\ \ref{fig:4}(a). The experimental
irreversibility line agrees rather well with the calculated melting curve. A
similar procedure is applied to $\kappa $-(ET)$_{2}$Cu(SCN)$_{2}$. The
parameter $H_{c2}\approx 4.7$~T was determined by fitting the additional
damping amplitude of the dHvA oscillation in the mixed SC state, as
calculated from our SC fluctuations theory \cite{RMP01}, to the
corresponding experimental data \cite{Sasaki98}. Again, the calculated
melting curve is found close to the irreversible line obtained in Ref.\ \cite
{Sasaki98} [Fig.\ \ref{fig:4}(b)]. It is interesting to note that at $T=0$
the resulting melting field coincides with the mean-field $H_{c2}$. The
sharp increase toward $H_{c2}$, characterizing $H_{m}$ at $T\rightarrow 0$,
is similar to that of $H_{irr}$ observed experimentally in both materials.

The relation $H_{irr}\approx H_{m}$ motivates us to exploit our
fluctuating-vortex-chain model above $H_{m}$ in calculating the
magnetic-field dependence of the jump $\Delta M_{\uparrow \downarrow }$
above $H_{irr}$, and compare it to the experimental result (Fig.\ \ref{Fig.1}%
). Combining Eq.\ (\ref{Ampere}) with Eq.\ (\ref{CritLorenz}), we may
rewrite the critical gradient as $\overline{\left| \partial B/\partial
x\right| }=4\pi \overline{\left| F_{\eta }\right| }/B=4\pi E_{fric}/L_{x}B$.
Since the generators of the friction force, i.e., the pinning force and the
vortex-vortex interactions, all originate in the existence of local SC order
at the vortex position, the friction energy $E_{fric}$ should be
proportional to the mean-square SC order parameter, $\langle |\Delta
_{0}(H)|^{2}\rangle $, and so the magnetization jump can be written as $%
\Delta M_{\uparrow \downarrow }=\frac{\zeta }{3\pi }\Delta B\approx \frac{%
L_{x}}{3\pi }\overline{\left| \frac{\partial B}{\partial x}\right| }\propto 
\frac{1}{B}\langle |\Delta _{0}(B)|^{2}\rangle $. Using the limit of
independently fluctuating vortex chains in the GL theory \cite{RMP01} to
describe the vortex-liquid state above $H_{irr}$, i.e., writing: $\langle
|\Delta _{0}(B)|^{2}\rangle =\sqrt{\frac{2k_{B}T}{\pi ^{2}\beta }}\Phi (\xi
) $, where $\Phi (\xi )=\sqrt{\pi }\Big[
\xi +\frac{\exp (-\xi ^{2})}{2\int_{-\infty }^{\xi }d\varsigma \exp
(-\varsigma ^{2})}\Big]$, we find that 
\begin{equation}
\Delta M_{\uparrow \downarrow }\propto \frac{1}{B}\langle |\Delta
_{0}(B)|^{2}\rangle \propto \Phi (\xi )  \label{DelMjump}
\end{equation}
is a universal function of the dimensionless parameter $\xi (B,T)$.
Employing the material parameters, $T_{c}=4.4$~K, $\frac{m_{c}}{m_{e}}=2.0$, 
$E_{F}/k_{B}=133$~K at the temperature of the experiment, $T=40$~mK, so that 
$\xi \approx 31(1-B/H_{c2})/B$, and treating $H_{c2}$ and the
proportionality factor in Eq.\ (\ref{DelMjump}) as adjustable parameters,
the best fit is obtained for $H_{c2}=2$~T (Fig.\ \ref{fig:3}). The resulting
curve reflects the crossover between the vortex-crystal state below $H_{c2}$%
, which is well described by mean-field theory, and the normal state far
above $H_{c2}$, with the long tail of the field-dependent magnetization
hysteresis corresponding to the enhanced influence of 2D SC fluctuations.

In conclusion, we have shown that the striking phenomenon of the weak
magnetization hysteresis loops, observed deep in the vortex-liquid state of
a nearly 2D superconductor can be reasonably explained as arising from shear
viscous flow of easily sliding vortex chains, which are clustered around a
small number of strong pinning centers.

This research was supported by a grant from the Israel Science Foundation
founded by the Academy of Sciences and Humanities, and by the fund from the
promotion of research at the Technion.

\end{document}